\documentclass[12pt, draftclsnofoot, onecolumn]{IEEEtran}
\usepackage{cite}
\usepackage[pdftex]{graphicx}
\usepackage{epstopdf}
\usepackage[cmex10]{amsmath}
\usepackage{algorithmic}
\usepackage{amssymb}
\usepackage{color}
\usepackage{bm}
\usepackage[keeplastbox]{flushend}
\usepackage{suffix}
\usepackage{mathtools}

\DeclarePairedDelimiterX\MeijerM[3]{\lparen}{\rparen}%
{\begin{smallmatrix}#1 \\ #2\end{smallmatrix}\delimsize\vert\,#3}

\newcommand\MeijerG[8][]{%
  G^{\,#2,#3}_{#4,#5}\MeijerM[#1]{#6}{#7}{#8}}

\WithSuffix\newcommand\MeijerG*[7]{%
  G^{\,#1,#2}_{#3,#4}\MeijerM*{#5}{#6}{#7}}

\begin{document}

\title{Performance of SWIPT-based Differential Amplify-and-Forward Relaying with Direct Link}

\author{Yi~Lou,
		Julian~Cheng,~\IEEEmembership{Senior~Member,~IEEE},
		Yan Zheng~and
        Hong-Lin~Zhao,~\IEEEmembership{Member,~IEEE}
\thanks{Y. Lou and H.-L. Zhao are with the Department of Communication Engineering, Harbin Institute of Technology, Harbin 150001, China (e-mail: hlzhao@hit.edu.cn).}
\thanks{J. Cheng is with the School of Engineering, The University of British Columbia, Kelowna, BC, V1X 1V7, Canada (e-mail:julian.cheng@ubc.ca)}
\thanks{Y. Zheng is with the Harbin Research Institute of Electrical Instruments, Harbin 150023, China (e-mail:hitzhengyan@gmail.com)}
\thanks{H.-L. Zhao is the corresponding author.}}

\maketitle

\begin{abstract}
A novel asymptotic closed-form probability density function (pdf) of the two-hop (TH) link is derived for a simultaneous wireless information and power transfer based differential amplify-and-forward system. Based on the pdf, asymptotic closed-form average bit-error rate expressions of the single TH link and the TH link with direct link combined with a linear combining scheme are both derived. Monte Carlo simulations verify the analytical expressions.
\end{abstract}

\section{Introduction}
\IEEEPARstart{S}{imultaneous} wireless information and power transfer (SWIPT) has been regarded as a promising solution to prolong the lifetime of energy-constrained nodes \cite{Zhang2013MIMO}. Meanwhile, cooperative relaying can improve the system performance and extend the coverage without additional infrastructure.

SWIPT-based relaying schemes, which embrace the benefits of SWIPT and relaying, have been studied in \cite{Liu2016Wireless,Liu2016Power,Tang2016Wireless,Zhang2016Secure,Zhang2016Beamforming,Liu2017A}.  Various
performance analyses have been conducted for SWIPT-based relaying systems in the literature. The authors in \cite{Nasir2013Relaying} derived the approximate analytical expressions for the ergodic capacity and the outage probability (OT) for delay-tolerant and delay-limited transmission modes of the two-hop (TH) amplify-and-forward (AF) system. The OT for decode-and-forward (DF) system employing relay selection scheme is derived in closed-form in \cite{Chen2016Cooperative}. For the AF system with direct link, the asymptotic closed-form OT of the maximum-ratio combining is obtained in closed-form in \cite{Lee2017Outage}.

The aforementioned works on SWIPT-based relaying systems have assumed the availability of the instantaneous channel state information (CSI) for the coherent decoding at the receiving nodes. To obtain the instantaneous CSI, channel estimation algorithms have to be implemented at the energy-constrained nodes, which increase the energy consumption, especially for the multi-relay systems. Moreover, the potential channel estimation error can also impact the system performance \cite{Seyfi2011Performance}. As a remedy, the SWIPT-based relaying schemes with differential modulation, which eliminate the need for channel estimation, have been analyzed in \cite{Liu2015Energy,Liu2015Noncoherent,Mohjazi2017Unified,Mohjazi2016Performance}. In particular, the authors in \cite{Liu2015Energy} have derived the exact maximum-likelihood detectors (MLDs) for power-splitting (PS) and time-switching (TS) differential DF (DDF) systems. Moreover, the closed-form approximate MLDs with lower complexity were also derived. For the PS and TS  differential AF (DAF) systems, the exact and the closed-form MLDs were also proposed by the same authors in \cite{Liu2015Noncoherent}. The authors in \cite{Mohjazi2017Unified} have analyzed the performance of the TH DAF system in terms of OT and average bit-error rate (ABER). However, both the OT and the ABER were both computed based on the numerical algorithm. Specifically, the OT was computed based on Euler numerical algorithm, and the ABER includes finite-range summation and integral involving integrands composed of approximations. The authors in \cite{Mohjazi2016Performance} presented a linear combining (LC) for the DAF system with direct link. However, the authors used the same derivation techniques to derive the ABER as that in the conventional DAF systems \cite{Zhao2005Performance,Zhao2007Differential,Zhao2008Performance}, and ignored the correlation between the SNR of each hop of the TH link, resulting from the energy harvesting operation. The derived ABER was valid only at very low SNR region and cannot provide insights for system design. To the best of our knowledge, the closed-form ABER for both the coherent and non-coherent SWIPT-based relaying with instantaneous CSI energy harvesting has not been investigated in prior work. 

In this letter, we study the SWIPT-based DAF system with direct link. Taking into account the correlation between the SNR of the source-relay link and the relay-destination link, a novel asymptotic closed-form probability density function (pdf) of the TH link is derived. Based on the pdf, simple and tight asymptotic expressions for the ABER of the single TH link and the TH link with direct link combined using the LC scheme are both derived in closed-form. Monte Carlo simulations verify the analytical expressions.

\section{System Model and Selection Combining}
As shown in Fig. 1, we consider a DAF relaying system where a source node S communicates with a destination node D with the help of an energy-constrained relay node R. All the nodes have a single antenna and work in half-duplex mode. Before transmission, the binary phase-shift keying modulated symbol $d[n] \in\{+1,-1\}$ is encoded differentially as $s[n]=s[n-1]d[n]$ with $s[0]=1$. The relaying protocol consists of two signal transmission phases, and the  duration of each phase is $T/2$. In the first phase, S transmits $s[n]$ to D and R with average power $P_0$. According to the PS protocol, the received signals at R are then assigned to the information processing (IP) module and the energy harvesting (EH) module. 

EH implementations can be classified into two types: instantaneous CSI EH (IEH)\cite{Liu2015Energy} and average CSI EH (AEH)\cite{Liu2015Noncoherent}. In AEH, each EH process lasts for a duration of multiple transmission blocks; therefore, the harvested energy is determined by the average CSI of the SR channel. In IEH, each EH process lasts for a duration of one transmission block, therefore, the harvested energy is determined by the instantaneous CSI of the SR channel. AEH usually requires a battery having sufficiently large storage capacity which is difficult to implement \cite{Liu2015Noncoherent}; In this work, we adopt the IEH scheme.

The channels are assumed to be Rayleigh block fading which remain unchanged during one time slot and change independently from one time slot to anther. The fading gains of the SD, SR, and RD links are modeled as $h_i[n]\sim\mathcal{CN}(0,1), i=0, 1, 2$, respectively, where $\mathcal{CN}(0,1)$ denotes a complex Gaussian distribution with mean 0 and variance 1 and $n=1,2,\dots$, denotes the symbol index. The distances for the SD, SR, and RD links are denoted as  $d_i, i=0, 1, 2$, respectively. In the PS protocol, the $\theta$ portion of the radio frequency (RF) received signals at R is harvested by EH module. The total harvested energy with IEH in phase I is $E_r=T P_0\eta\theta|h _1|^2/(2 d_1^{\alpha})$, where $\alpha$ is the path loss exponent and $\eta$ is the energy conversion efficiency. The harvested power $P_r=2E_r/T=P_0 \eta \theta/d_1^{\alpha }$. The IP module of R down-converts the remaining $\phi=(1-\theta)$ portion of the RF received signals to baseband. The received signals at D and the IP module of R are, respectively, given by
\begin{figure}[!t]
\centering
\includegraphics[width=3.45in]{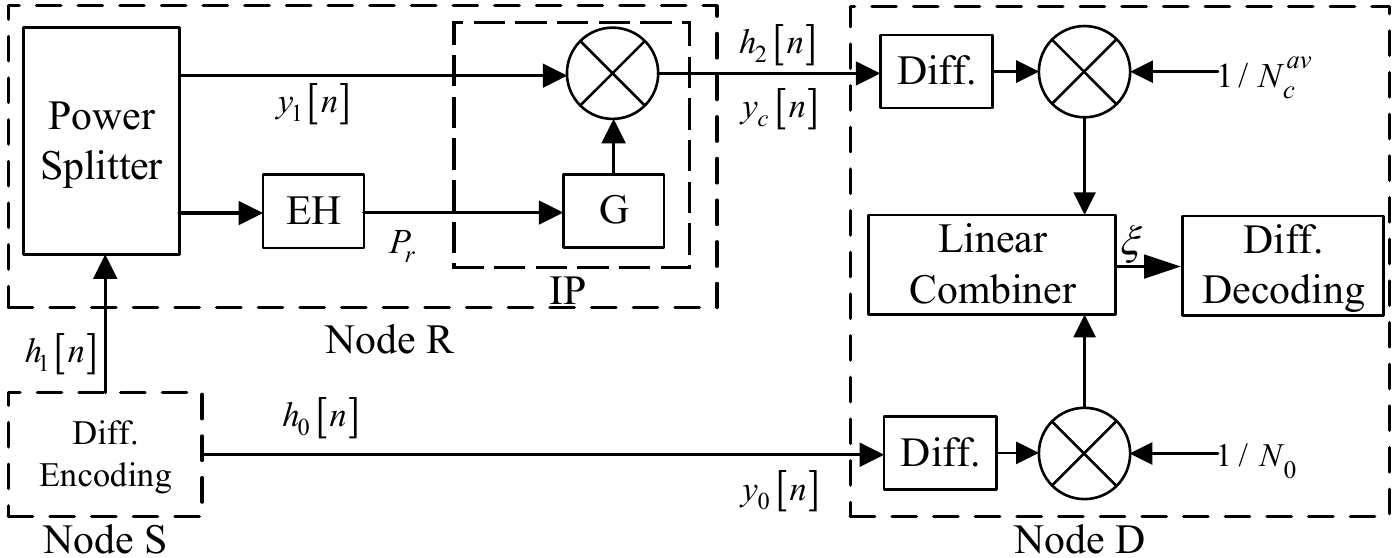}
\caption{Block diagram of a three-node SWIPT-based DAF system model with direct link using the LC scheme.}
\label{model}
\end{figure}
\begin{IEEEeqnarray}{lCl}
y_0[n]&=&\frac{\sqrt{P_0} h_0[n] s[n]}{\sqrt{d_0^{\alpha}}}+w_{0}[n],\\
y_1[n]&=&\frac{\sqrt{P_0} \sqrt{\phi} h_1[n] s[n]}{\sqrt{d_1^{\alpha}}}+w_1[n].
\end{IEEEeqnarray}

 For $i=0,2$, $w_i[n]=w_{ia}[n]+w_{ic}[n]$ are the overall Gaussian noises with variance $N_i=N_{ia}+N_{ic}$ at D in phase I and phase II, respectively. Whereas $w_1[n]=\sqrt{\phi} w_{1a}[n]+w_{1c}[n]$ is the overall Gaussian noises with variance $N_1=\phi N_{1a}+N_{1c}$ at the IP module of R. For $i=0,1,2$, $w_{ia}$ and $w_{ic}$ are the Gaussian noise with variance $N_{ia}$ and $N_{ic}$, due to the receive antenna and the down-conversion, respectively. Obviously, the SNR of the direct link is
\begin{IEEEeqnarray}{lCl}\label{gammad}
\gamma _0=\frac{P_0 d_0^{-\alpha}}{N_0}.
\end{IEEEeqnarray}

In Phase II, the received signals at R is scaled by an amplification factor $G$ to normalized the forwarded power per symbol from R to $P_r$. Based on the harvested power $P_r$ and the path loss of the SR link $d_1^\alpha$, the fixed-gain factor is given by
\begin{IEEEeqnarray}{C}\label{amplify gain}
G=\sqrt{\frac{P_r}{\mathbb{E}\{|y_1[n]|^2\}}}=\sqrt{\frac{P_r}{N_{1} +d_1^{-\alpha } P_0 \phi }}.
\end{IEEEeqnarray}

The corresponding received signals at D can be expressed as
\begin{IEEEeqnarray*}{lCl}\label{yc}
y_c[n]&=& \frac{G y_1[n]h_2[n]}{\sqrt{d_2^{\alpha}}}+w_2[n]\\
&=&\left(d_1 d_2\right)^{-\alpha /2} P_0 \sqrt{\frac{\eta  \phi  \theta }{d_1^{\alpha } N_1+P_0 \phi }} \left|h_1[n]\right| h_1[n] h_2[n] s[n]\\
&&+\underbrace{\sqrt{\frac{P_0\eta  \theta }{\left(d_1 d_2\right)^{\alpha } N_1+d_2^{\alpha } P_0 \phi }} |h_1[n]|h_2[n] w_1[n]+w_2[n]}_{w_c}\\\IEEEyesnumber
\end{IEEEeqnarray*}
where $w_c[n]$ is the equivalent noise of the TH link. After some simplification, the variance $N_c$ of the equivalent noise $w_c[n]$ can be derived as
\begin{IEEEeqnarray}{lCl}\label{nc}
N_c&=&N_2+\frac{N_1 P_0 \eta  \theta  \left|h_1[n]|^2\right|h_2[n]|^2}{\left(d_1 d_2\right)^{\alpha } N_1+d_2^{\alpha } P_0 \phi }.
\end{IEEEeqnarray}

To achieve the cooperative diversity, the received signals from S in Phase I and that from R in Phase II are combined using the LC scheme, and the output of the linear combiner is computed as
\begin{IEEEeqnarray}{lCl}
\xi=\frac{1}{N_0} y_0[n]y^*_0[n-1]+\frac{1}{N_c^{av}} y_c[n]y^*_c[n-1]
\end{IEEEeqnarray}
where $N_c^{av}$ is the mean value of $N_c$, and is obtained by replacing $|h_1[n]|^2|h_2[n]|^2$ in \eqref{nc} with the mean value 1 \cite{Patel2006Statistical}. The reason of substituting $N_c$ with $N_c^{av}$ is that the instantaneous CSIs are not available at D for the relaying systems using differential modulation.

Finally, the information symbol is detected as
\begin{IEEEeqnarray}{lCl}\label{relaysnr}
\hat{d}[n]=\text{sign} \left\{\xi \right\}.
\end{IEEEeqnarray}

\section{Performance Analysis}
In this section, we first derive the exact pdf of TH link involving a single integral, then a novel asymptotic pdf is derived in a closed-form. Based on the asymptotic pdf, the asymptotic ABERs of the single TH link and  TH link with direct link combined with LC are both derived in a tight and closed-form. 

With \eqref{yc} and after some simplification, the SNR of the TH link can be expressed as
\begin{IEEEeqnarray}{lCl}\label{gammac}
\gamma _c=\frac{M}{N}=\frac{k_1 X^2 Y}{k_3+k_2 X Y}
\end{IEEEeqnarray}
where
\begin{IEEEeqnarray*}{lCl}
k_1&=&d_1^{-\alpha } P_0^2 \eta  \theta  \phi,\label{k1}\IEEEyesnumber\\
k_2&=&N_1 P_0 \eta \theta,\label{k2}\IEEEyesnumber\\
k_3&=&d_2^{\alpha }\left(d_1^{\alpha }N_1 N_2+ N_2p_s \phi \right)\label{k3}
\end{IEEEeqnarray*}
and $X=|h_1[n]|^2$, $Y=|h_2[n]|^2$. Since $X$ and $Y$ are independent exponential random variables (RV) with parameter 1, it is easy to show that
\begin{IEEEeqnarray}{lCl}\label{pdfxy}
f_{X,Y}(x,y)&=&e^{-x-y},\quad x>0\&y>0.
\end{IEEEeqnarray}

Due to the common terms $X$ and $Y$,  random variables $M$ and $N$ are correlated. From \eqref{gammac}, it can be easily shown that
\begin{IEEEeqnarray*}{lCl}
X&=&-\frac{k_2 M}{k_1 \left(k_3-N\right)}\label{expression x},\IEEEyesnumber\\
Y&=&\frac{k_1 \left(k_3-N\right)^2}{k_2^2 M}.
\end{IEEEeqnarray*}

Hence, the Jacobian determinant of the transformation from $(X,Y)$ to $(M,N)$ can be computed as
\begin{IEEEeqnarray*}{lCl}
D_J&=&|J(m(x,y),n(x,y))|=\left|\begin{array}{cc} \partial _xm(x,y) & \partial _xn(x,y) \\ \partial _ym(x,y) & \partial _yn(x,y) \\\end{array}\right|\\
&=&\frac{1}{m N_1 P_0 \eta  \theta }.\IEEEyesnumber
\end{IEEEeqnarray*}

Since the transformation is invertible, after substituting $k_1$ and $k_2$ in \eqref{k1} and \eqref{k2} we have
\begin{IEEEeqnarray*}{lCl}
f_{M,N}(m,n)&=&D_Jf_{X,Y}\\
&&\times\left(-\frac{d_1^{\alpha } m N_1}{\left(k_3-n\right) P_0 \phi },\frac{d_1^{-\alpha } \left(k_3-n\right)^2 \phi }{mN_1^2 \eta  \theta }\right).\quad\IEEEyesnumber
\end{IEEEeqnarray*} 

From \eqref{pdfxy} and \eqref{expression x}, it is easy to show that $n>k_3$. With the help of \cite[eq. (6.60)]{Papoulis2002Probability}, the pdf of the TH link $f_{\gamma _c}(z)$ can be expressed as
\begin{IEEEeqnarray*}{lCl}\label{pdfgammac}
f_{\gamma _c}(z)&=&\int _{k_3}^{\infty }n f_{M,N}(n z,n)dn\\
&=&\int _0^{\infty }\left(t+k_3\right) f_{M,N}\left(\left(t+k_3\right) z,\left(t+k_3\right)\right)dt\\
&=&\int_0^{\infty } \frac{e^{-\frac{d_1^{\alpha } \left(k_3+t\right) N_1 z}{t P_0 \phi }-\frac{d_1^{-\alpha } t^2 \phi }{\left(k_3+t\right) N_1^2 z\eta  \theta }}}{N_1 P_0 z \eta  \theta } \, dt.\IEEEyesnumber
\end{IEEEeqnarray*} 

To the best of our knowledge, the integration in \eqref{pdfgammac} cannot be further simplified. Note that, the pdf in \eqref{pdfgammac} is different \cite[eq. (10)]{Mohjazi2016Performance}, where the authors assumed the RVs $M$ and $N$ are independent and therefore significantly simplified the derivation. However, from \eqref{gammac}, we know that the RVs $M$ and $N$ are correlated. Moreover, in Section IV, we can see that derived ABERs in \cite{Mohjazi2016Performance} were valid only at very low SNR region.

With the pdf in \eqref{pdfgammac}, the exact ABERs of the single TH link and the TH link with direct link combined with LC can be derived \cite{Simon2005Digital}, respectively, as
\begin{IEEEeqnarray}{lCl}
p_e^{\text{TH}}&=&\int _0^{\infty }\frac{1}{2}e^{-z}f_{\gamma _c}(z)dz,\label{ABERRENUM}\\
p_e^{\text{LC}}&=&\int _0^{\infty }\int _0^{\infty }\frac{1}{8}(4+z+x)e^{-z-x} f_{\gamma_0}(x)f_{\gamma_c}(z)dz dx\label{ABERLCNUM}\quad\ 
\end{IEEEeqnarray} 
where $f_{\gamma_0}(x)$ is the pdf of $\gamma_0$, and is expressed as
\begin{IEEEeqnarray}{lCl}
f_{\gamma_0}(x)&=& \frac{1}{\bar{\gamma }_0}e^{-\frac{x}{\bar{\gamma }_0}}
\end{IEEEeqnarray} 

The integrals in \eqref{ABERRENUM} and \eqref{ABERLCNUM} prevent us from obtaining helpful insights on the system design. In what follows,
we derive a simple and closed-form pdf of the TH link based on a high SNR approximation. At high SNR, $t$ in the denominator of the second fraction in exponent in \eqref{pdfgammac} is negligible, compared to $k_3$, since $k_3$ is a function of $P_0$. By setting the $t$ to 0, the $f_{\gamma _c}(z)$ in \eqref{pdfgammac} can be approximated as
\begin{IEEEeqnarray*}{lCl}\label{apprpdf}
f_{\gamma _c}^*(z)&=&\int_0^{\infty } \frac{e^{-\frac{d_1^{\alpha } \left(k_3+t\right) N_1 z}{t P_0 \phi }-\frac{d_1^{-\alpha } t^2 \phi }{k_3 N_1^2z \eta  \theta }}}{N_1 P_0 z \eta  \theta } \, dt\\
&=&\frac{e^{-\frac{d_1^{\alpha } N_1 z}{P_0 \phi }}}{N_1 P_0 z \eta  \theta }\int _0^{\infty }e^{-\frac{d_1^{\alpha } k_3 N_1 z}{P_0 t \phi }}e^{-\frac{d_1^{-\alpha} t^2 \phi }{k_3 N_1^2 z \eta  \theta }}dt\IEEEyesnumber.
\end{IEEEeqnarray*} 

With \cite{MMA} (01.03.26.0004.01) and (07.34.17.0012.01), we have the following results
\begin{IEEEeqnarray*}{lCl}
e^{-\frac{d_1^{\alpha } k_3 N_1 z}{P_0 t \phi }}&=&G_{1,0}^{0,1}\left(\left.\frac{d_1^{-\alpha } P_0 t \phi }{k_3 N_1 z}\right\rvert\arraycolsep=0.1pt\begin{array}{c}1 \\-\end{array}\right),\\
e^{-\frac{d_1^{-\alpha } t^2 \phi }{k_3 N_1^2 z \eta  \theta }}&=&G_{0,1}^{1,0}\left(\left.\frac{d_1^{-\alpha } t^2 \phi }{k_3 N_1^2 z \eta  \theta }\right\rvert\arraycolsep=0.1pt\begin{array}{c} -\\0 \\\end{array}\right)
\end{IEEEeqnarray*} 
where $G_{p,q}^{m,n}(x)$ is the Meijer G-function \cite[eq. (9.301)]{Gradshteyn2007Table}, which is available as a standard built-in function in the most of popular mathematical software packages, such as Maple, MATLAB,  and Mathematica. Using Mellin transform of the product of two Meijer G-functions \cite[eq. (3.21)]{Mathai1973Generalized}, the integral in \eqref{apprpdf} can be derived as
\begin{IEEEeqnarray*}{lCl}\label{pdfapprresult}
f_{\gamma _c}^*(z)&=&\frac{e^{-\frac{d_1^{\alpha } N_1 z}{P_0 \phi }}}{N_1 P_0 z \eta  \theta }\int _0^{\infty }G_{1,0}^{0,1}\left(\left.\frac{d_1^{-\alpha} P_0 t \phi }{k_3 N_1 z}\right\rvert\arraycolsep=0.1pt\begin{array}{c} 1\\- \\\end{array}\right)\\
&&\times G_{0,1}^{1,0}\left(\left.\frac{d_1^{-\alpha } t^2 \phi }{k_3 N_1^2 z \eta  \theta }\right\rvert\arraycolsep=0.1pt\begin{array}{c} -\\0 \\\end{array}\right)dt\\
&=&\frac{e^{-\frac{d_1^{\alpha } N_1 z}{P_0 \phi }} j_1 }{\sqrt{\pi }}G_{0,3}^{3,0}\left(j_1 z\left\rvert\arraycolsep=0.1pt\begin{array}{c} -\\-\frac{1}{2},0,0 \\\end{array}\right.\right)\IEEEyesnumber
\end{IEEEeqnarray*} 
where $j_1=\frac{d_1^{\alpha } k_3}{4 P_0^2 \eta  \theta  \phi}$. Notably, this is the first closed-form asymptotic expression of the pdf of the TH link for SWIPT-based DAF systems, and verified to be very tight for the whole region of $P_0$ in section IV.

By substituting $f_{\gamma _c}(z)$ in \eqref{ABERRENUM} with $f_{\gamma _c}^*(z)$ in \eqref{pdfapprresult}, and with the help of \cite{MMA} (01.03.26.0004.01) and \cite[eq. (7.811)]{Gradshteyn2007Table}, the asymptotic closed-form ABER of the single TH link can be derived as
\begin{IEEEeqnarray*}{lCl}\label{aberthappr}
p_e^{\text{TH}^*}=\int _0^{\infty }f_{\gamma _c}^*(z) *\frac{1}{2}e^{-z}dz=\frac{1}{j_2}G_{1,3}^{3,1}\left(\left.\frac{j_1}{j_2}\right\rvert\arraycolsep=0.1pt\begin{array}{c} 0 \\ -\frac{1}{2},0,0 \\\end{array}\right)\quad\IEEEyesnumber
\end{IEEEeqnarray*} 
where $j_2=-1-\frac{d_1^{\alpha } N_1}{P_0 \phi }$.

By substituting $f_{\gamma _c}(z)$ in \eqref{ABERLCNUM} with $f_{\gamma _c}^*(z)$ in \eqref{pdfapprresult} and after some manipulations, we have
\begin{IEEEeqnarray*}{lCl}\label{}
p_e^{\text{LC}^*}&=&\frac{1}{8\bar{\gamma }_0}\int _0^{\infty }e^{-\frac{x}{\bar{\gamma }_0}-x} dx\bigg(\underbrace{\int _0^{\infty }(4+x)e^{-z} f_{\gamma _c}^*(z)dz}_{I_1}\\
&&+\underbrace{\int_0^{\infty }z e^{-z} f_{\gamma _c}^*(z)dz}_{I_2}\bigg).\IEEEyesnumber
\end{IEEEeqnarray*} 

Using \cite{MMA} (07.34.17.0011.01), we have
\begin{IEEEeqnarray*}{lCl}
z e^{-z}&=&G_{0,1}^{1,0}\left(z\left\rvert\arraycolsep=0.1pt\begin{array}{c} -\\1 \\\end{array}\right.\right).
\end{IEEEeqnarray*} 

Follow a similar approach in \eqref{aberthappr}, the first integral $I_1$ and the second integral $I_2$ can be derived, respectively, as
\begin{IEEEeqnarray}{lCl}
I_1&=&\frac{2(x+4)}{j_2}G_{1,3}^{3,1}\left(\left.\frac{j_1}{j_2}\right\rvert\arraycolsep=0.1pt\begin{array}{c} 0 \\ -\frac{1}{2},0,0 \\\end{array}\right),\label{i1}\\
I_2&=&\frac{1}{j_2 \sqrt{\pi }}G_{1,3}^{3,1}\left(\frac{j_1}{j_2}\left\rvert\arraycolsep=0.1pt\begin{array}{c} 0 \\ \frac{1}{2},1,1 \\\end{array}\right.\right).\label{i2}
\end{IEEEeqnarray}

With \eqref{aberthappr}, \eqref{i1} and \eqref{i2}, the asymptotic ABER of the LC scheme can be derived in a closed-form as
\begin{IEEEeqnarray*}{lCl}\label{aberapprlc}
p_e^{\text{LC}^*}&=&\frac{j_1 \left(4+5 \bar{\gamma }_0\right)}{8 j_2 \sqrt{\pi } \left(1+\bar{\gamma }_0\right)^2}G_{1,3}^{3,1}\left(\frac{j_1}{j_2}\left\rvert\arraycolsep=0.1pt\begin{array}{c} 0 \\ -\frac{1}{2},0,0 \\\end{array}\right.\right)\\
&&+\frac{1}{8 j_2 \sqrt{\pi } \left(1+\bar{\gamma }_0\right)}G_{1,3}^{3,1}\left(\frac{j_1}{j_2}\left\rvert\arraycolsep=0.1pt\begin{array}{c} 0 \\ \frac{1}{2},1,1 \\\end{array}\right.\right).\IEEEyesnumber
\end{IEEEeqnarray*}

\section{Simulation Results}
In this section, Monte Carlo simulations are presented to verify the derived asymptotic closed-form ABER expressions and to provide insights into the impact of various parameters on the system performance. For simplicity, all the variances of the down-conversion noise the receive antenna noise are set to $\frac{1}{2}$, i.e., $N_{ia}=N_{ic}=\frac{1}{2}$, for $i=0,1,2$. 

Figure 2 plots the exact ABERs in \eqref{ABERRENUM} and \eqref{ABERLCNUM}, and the asymptotic closed-form ABERs in \eqref{aberthappr} and \eqref{aberapprlc}, for TH scene with label IEH-TH and LC scene with label IEH-LC, respectively. The analytical ABERs in (11) and (12) in \cite{Mohjazi2016Performance}, for TH and LC scenes, respectively, are also plotted for comparative purposes. Moreover, the conventional (CON) DAF system \cite{Zhao2005Performance,Zhao2007Differential,Zhao2008Performance} and the AEH-based DAF system \cite{Liu2015Noncoherent} are also plotted. Firstly, Figure 2 shows that the analyses in \cite{Mohjazi2016Performance} have severe deviations from the simulation results for both TH and LC scenes. The reason is that the authors in \cite{Mohjazi2016Performance} used the same derivation techniques to derive the ABER as that in the CON DAF system. However, as shown in \eqref{gammac}, the SNR of the TH link $\gamma_c$ is more complicated for the SWIPT-based DAF system compared with that in the CON DAF system, resulting from the energy harvesting operation. Secondly, the derived exact ABERs agree well with simulations for both scenes. Thirdly, the derived asymptotic closed-form ABERs are close to the simulations at the small $P_0$ region and match perfectly with simulations for $P_0>15\ \textrm{dB}$ for both scenes. Since one additional CSI of SR link $h_1$ is induced into the received signals, the channels in IEH-based schemes can be regarded as a $3\times\textrm{Rayleigh}$ cascaded channels, resulting in an inferior performance than that of the CON and AEH-based schemes \cite{Dohler2010Cooperative}.

Figure 3 plots the ABERs for IEH-based DAF systems in TH and LC scenes for different values of the SR distance, $d_1$. The RD distance, $d_2$ is set to $d_2=3-d_1$. One can see that the asymptotic ABERs are very tight for all the scenes. For each value of $\theta$, the ABER decreases for $d_1<1.5$. This is because increasing $d_1$ means increasing path loss $d_1^\alpha$; therefore, both the received signal strength at the IP module of R and the harvested energy by the EH module at R decrease. For each value of $\theta$, the ABER increases for $d_1>1.5$. This is because a larger $d_1$ means a smaller path loss $d_1^\alpha$, and hence even lesser transmitted power at R is sufficient to support the reliable communication from R to D. Therefore, the position of R with the worst error performance is the midpoint of the SR link. Notably, this is different from that of the CON DAF relaying system where the midpoint is the optimal location of R. 

\begin{figure}[!t]
\centering
\includegraphics[width=3.45in]{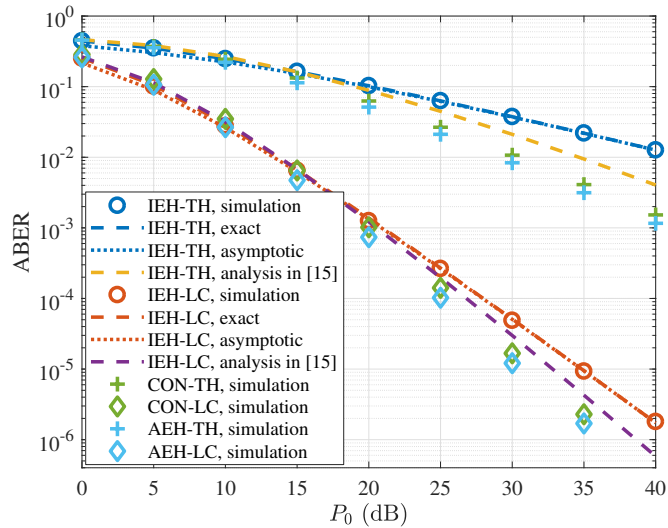}
\caption{ABER performance of the IEH, CON, and AEH-based DAF systems in TH and LC scenes with $d_0=d_1=d_2=1$, $\eta=0.7$, $\theta=0.5$ and $\alpha=2$. }
\label{ber}
\end{figure}
\begin{figure}[!t]
\centering
\includegraphics[width=3.45in]{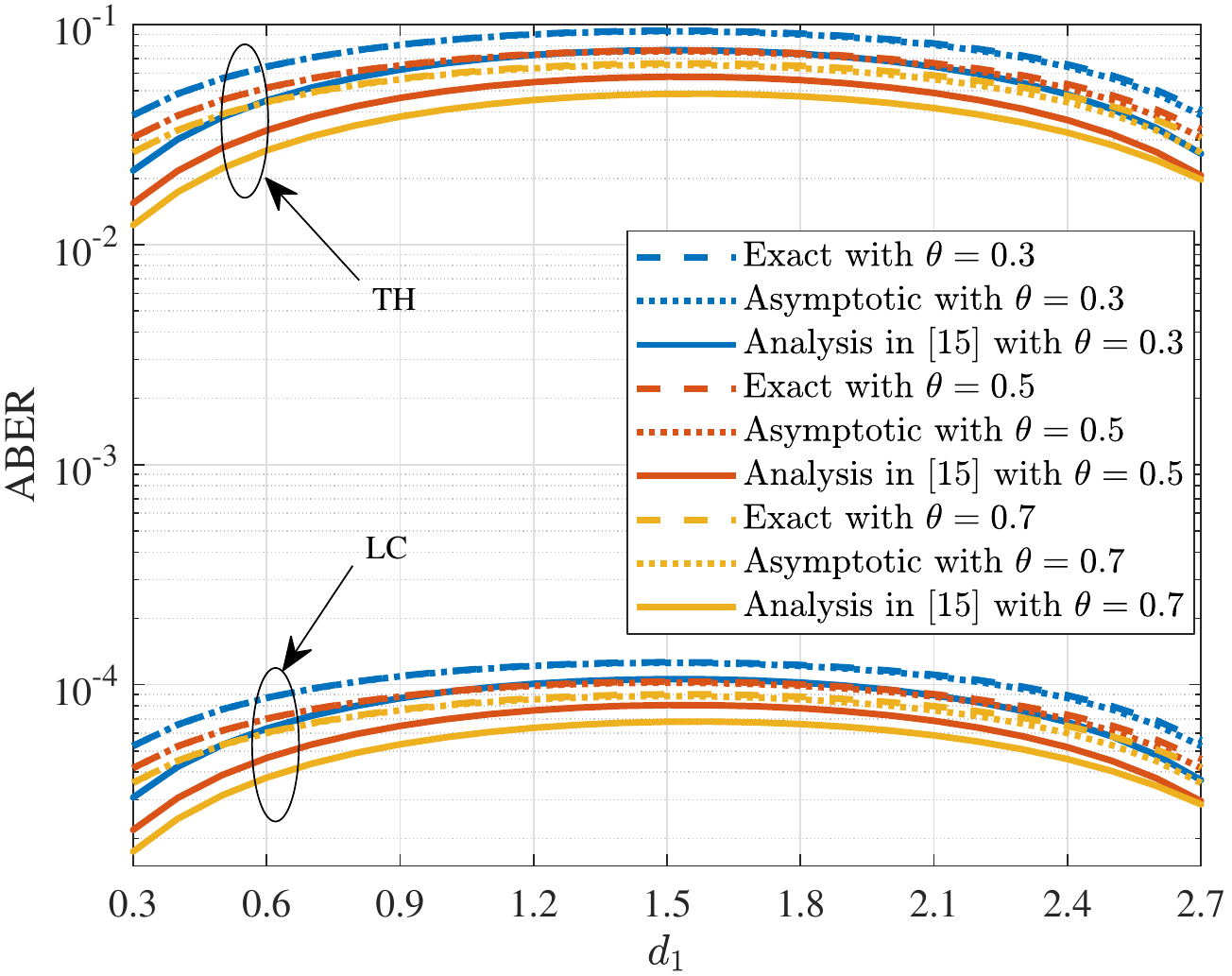}
\caption{ABER performance of the IEH-based DAF systems in TH and LC scenes with $d_0=1$, $d_2=3-d_1$, $\eta=0.7$ and $\alpha=2$. }
\label{ber}
\end{figure}

\section{Conclusion}
In this letter, we studied the performance of a three nodes SWITP-enabled differential amplify-and-forward relaying system with direct link, employing linear combiner at the destination node. A novel asymptotic probability density function of the two-hop link is derived in closed-form. Based on the pdf, simple and tight asymptotic expression for the ABER of the linear combining scheme is derived in closed-form. The impact of various parameters
on the system performance are also investigated.

\bibliographystyle{IEEEtran}
\bibliography{IEEEabrv,ref}
\end{document}